# P/2019 LD2 (ATLAS): An Active Centaur in Imminent Transition to the Jupiter Family

J.K. Steckloff[1,2], G. Sarid[3], K. Volk[4], T. Kareta[4], M. Womack[5], W. Harris[4], L. Woodney[6], C. Schambeau[5]

[1]Planetary Science institute, Tucson, AZ
[2]University of Texas at Austin, Austin, TX
[3]SETI Institute, Mountain View, CA
[4]Lunar and Planetary Laboratory, University of Arizona, Tucson, AZ
[5]Florida Space Institute and Department of Physics, University of Central Florida, Orlando, FL
[6]California State University, San Bernardino, CA

## Abstract

The recently discovered object P/2019 LD2 (ATLAS) was initially thought to be a Jupiter Trojan asteroid, until dynamical studies and the appearance of persistent cometary activity revealed that this object is actually an active Centaur. However, the dynamical history, thermal environment, and impact of such environments on the activity of 2019 LD2 are poorly understood. Here we conduct dynamical simulations to constrain its orbital history and resulting thermal environment over the past 3000 years. We find that 2019 LD2 is currently in the vicinity of a dynamical "Gateway" that facilitates the majority of transitions from the Centaur population into the Jupiter Family of Comets (JFC population; Sarid et al. 2019). Our calculations show that it is unlikely to have spent significant amounts of time in the inner solar system, suggesting that its nucleus is relatively pristine in terms of physical, chemical, and thermal processing through its history. This could explain its relatively high level of distant activity as a recently activated primordial body. Finally, we find that the median frequency of transition from the Gateway population into the JFC population varies from once every ~3 years to less than once every 70 years, if 2019 LD2's nucleus is ~1 km in radius or greater than 3 km in radius. Forward modeling of 2019 LD2 shows that it will transition into the JFC population in 2063, representing the first known opportunity to observe the evolution of an active Centaur nucleus as it experiences this population-defining transition.

1. **Introduction**

The Centaurs are a dynamically unstable population of icy bodies orbiting the Sun within the region of the giant planets (between ~5 and 30 au from the Sun). These objects originate from their trans-Neptunian reservoir population (chiefly the dynamically excited Scattered Disk and hot classical Kuiper Belt) via gravitational interactions with Neptune, which slowly feeds objects into the Centaur population over ~1 Gyr timescales (e.g., Duncan & Levison 1997; Duncan et al. 2004; Dones et al. 2015). Centaurs subsequently evolve dynamically through gravitational interactions with the giant planets over ~1-10 Myr timescales, until they are either ejected from the solar system or migrate into the inner solar system (e.g., Tiscareno & Malhotra 2004; Di Sisto & Brunini 2007; Sarid et al. 2019), where they are reclassified as Jupiter Family Comets (JFCs).

Because many common cometary volatiles become warm enough to drive activity in the Centaur region (Jewitt 2009; Steckloff & Jacobson 2016; Womack et al. 2017; Safrit et al. 2020), these objects sometimes appear as bare asteroidal bodies and other times as active, cometary bodies. This activity can be profound, forming comae (see, e.g., Meech & Belton 1990; Bauer et al. 2008; Jewitt 2009; Seccull et al. 2019), ejecting fragments and debris (Bauer et al. 2008; Rousselot 2008; Kareta et al. 2019), and spinning up nuclei into bilobate shapes (Safrit et al. 2020). Centaurs experience negligible collisional evolution (Durda & Stern 2000), and rarely approach sufficiently close to the giant planets to experience tidal deformation/disruption (Safrit et al. 2020; c.f., Hyodo et al. 2016), Nevertheless such deep encounters do occasionally happen, as evidenced by comet Shoemaker-Levy 9's disruption (Marsden 1993; Asphaug & Benz 1994; Chodas & Yeomans 1996). In short, the physical evolution of Centaurs is likely dominated by outbursts and other volatile production mechanisms. This makes them ideal targets for studying the isolated effects of these thermodynamic processes on small icy bodies.

Recent forward modeling of TNOs through the giant planet region found that there is a specific dynamical pathway that facilitates the transition between the Centaur and JFC populations (Sarid et al. 2019). In particular, the majority of objects that eventually evolve into JFCs leave the Centaur population through a dynamical "Gateway," and objects in Gateway orbits are likely to transition into JFCs in the near future (Sarid et al. 2019). However, bright, highly active objects in Gateway orbits are unlikely to have spent significant prior time as JFCs, despite the dynamics of the Gateway allowing for reverse transitions to occur. JFC nuclei that would have evolved outward to Centaur-like orbits, would have also experienced significant fading (Sarid et al. 2019; Brasser & Wang 2015). Thus, active objects currently in the Gateway region represent particularly compelling targets to investigate how dynamical and thermodynamic evolution alters primordial objects prior to becoming a JFC.

Sarid et al. (2019) identified four Centaurs[1] currently residing in the orbital Gateway: P/2010 TO20 (LINEAR-Grauer), P/2008 CL94, 2016 LN8, and 29P/Schwassmann-Wachmann (hereafter SW1). Of these identified objects, dynamical studies (combined with fading laws suggest that the highly active Centaur SW1 is very likely to exit the Gateway and transition into the JFC population within approximately 10,000 years (Sarid et al. 2019). This transition will present future residents of Earth with a short-period comet with the potential to rival or exceed the activity and brightness of the great Comet Hale-Bopp!

In 2019, the ATLAS project discovered a new object, P/2019 LD2 (ATLAS) (hereafter, "LD2") along an orbital arc that initially suggested it was a Jupiter Trojan asteroid. However, additional observations and dynamical analysis revealed it to be in an unstable orbit, a result of a close encounter with Jupiter in 2017 (Kareta et al. 2020; Hsieh et al. 2020). Comet-like activity was suspected in the discovery frames (MPEC 2020-K134[2]) and subsequently confirmed in images throughout summer 2019. Recovery in 2020 by ATLAS[2] showed that the object was persistently active with a coma of diameter 8-12 arcsec, slight anti-sunward tail elongation up to 30 arsec, and with an apparent magnitude of 17.6-18.5. These observations argue against intermittent causes for LD2's activity such as impacts (Durda & Stern 2000), mass wasting (Steckloff et al. 2016; Steckloff & Samarasinha 2018), and radiative (Bottke et al. 2006) or sublimative spin up (Steckloff & Jacobson 2016; Safrit et al. 2020), in favor of volatile sublimation or other thermally driven processes (as similarly argued in Hsieh et al. 2004 for the activity of Main-Belt Comets).

Together, these observed dynamics and activity show that LD2 is an active Centaur in a quickly evolving transition orbit. Moreover, its orbit is in the vicinity of the dynamical Gateway region (Sarid et al. 2019), and will likely transition into a JFC in ~40 years (Kareta et al. 2020; Hsieh et al. 2020). This makes LD2 a particularly compelling object to investigate while it is still a Centaur, as observational studies in the future will reveal how it responds to the intense thermal environment experienced by JFCs. Considering that the short-term dynamical history of LD2 has already been studied (Kareta et al., 2020, Hsieh et al., 2020), an investigation into its long-term orbital history is critical to understanding the object's current activity relates to its past and future evolution. In particular, such an analysis is crucial to understanding whether LD2 is 'pristine', that is, preparing to enter the inner Solar System for the first time.

## 2. Methods

---

[1] Some more objects may still be identified within small body catalogues, as orbits in close proximity to the Gateway tend to either vary on a relatively short timescale or include correction to orbital elements with more observations.
[2] https://www.minorplanetcenter.net/mpec/K20/K20KD4.html
[2] http://www.ifa.hawaii.edu/info/press-releases/2019LD2/

We use a combination of dynamical and thermal evolution modeling to investigate the recent dynamical history of LD2. Centaur orbits are inherently chaotic (see, e.g., Tiscareno & Malhotra 2003) and can only be integrated forward/backward in time deterministically until relatively close encounters with the giant planets occur, after which dynamical integrations become highly sensitive to accumulated errors and uncertainties in the object's orbit. However, we can achieve a probabilistic description of the LD2's orbital evolution over limited timescales through dynamical backward integration of orbital clones. Even this probabilistic approach breaks down over longer timescales as the clones become distributed throughout phase space (the arrow of time becomes indistinct; see, e.g., a full discussion of this in Morbidelli et al. 2020) and results cease to become meaningful. Understanding these limits is therefore essential in interpreting backward integrations that enter this chaotic regime.

We use backwards numerical integrations to gain insight into LD2's recent orbital evolution (and resulting thermal history) over the past 3000 years. We choose this timescale based on forward modeling of the Centaur to JFC transition in Sarid et al. 2019. Our integrations here cover approximately twice the typical timescale that a Centaur will spend in the Gateway orbital region just outside Jupiter before transitioning to a JFC orbit. . As we discuss below, this timescale is longer than the period over which LD2's orbit can be deterministically followed into the past due to close encounters with Jupiter and Saturn; however, when placed in the context of our previous forward modeling results (Sarid et al. 2019), we are reasonably confident that our simulation results over this timescale represent a reasonable statistical sampling of LD2's possible past histories. Following Kareta et al. 2020, we integrate the orbit using the adaptive stepsize IAS15 integrator (Rein & Spiegel 2015) within REBOUND (Rein & Liu 2012). We generated 1000 clones of LD2 by sampling the JPL orbit fit covariance matrix (JPL Horizons Orbit Solution dated 2020 May 20 00:43:28,epoch JD 2458457.5). We integrate these clones 3000 years backward in time with a maximum timestep of 0.01 years, outputting the orbital state of each clone every 2.5 years.

We then examine these evolving orbits to compute how long each clone spends in specified heliocentric distance bins (of 0.5 au resolution) to constrain the likely evolution of LD2's recent thermal environment. We use these distance distributions to compute the magnitude of solar heating during this dynamical evolution and understand its evolving thermal environment, which is likely driving its activity and changes in its distribution of volatiles.

### 3. Results and Discussion

As discussed in Kareta et al. (2020), LD2 entered its current near-Jupiter orbit following a close encounter with Jupiter in 2017 (see below). Prior to that, it had an orbit in close proximity to the Gateway region, which allows for relatively strong interactions with Jupiter at perihelion and Saturn at aphelion. This behavior is characteristic of Centaurs on the cusp of entering the inner solar system (inside of ~4 au; Sarid et al 2019). LD2's very recent orbital history is quite deterministic until the 1770s, beyond which interactions with Jupiter and Saturn makes further backward integration highly chaotic, as evidenced in Figure 1 by the marked divergence of the clones' evolution. This divergence is the result of small differences between the clones' positions and velocities during a deep encounter with Jupiter (at a distance of ~0.3 au) in the 1850s, which leads to strong divergence of the clones's calculated pre-encounter orbital histories.

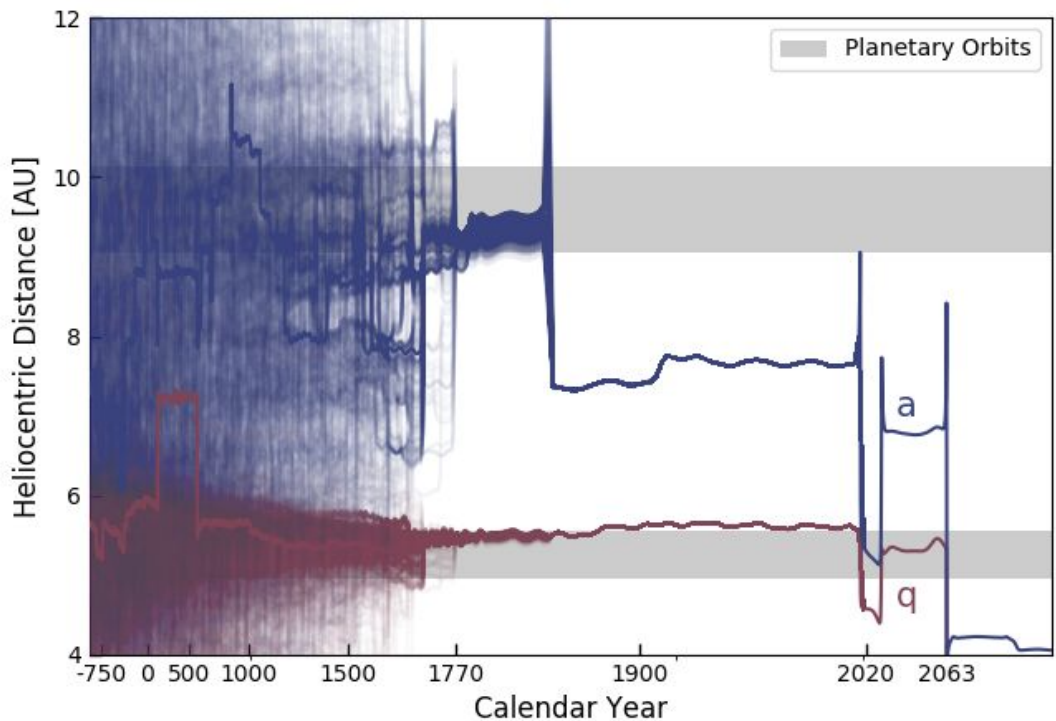

**Figure 1.** The backwards orbital evolution of the nominal orbit and 1000 clones of 2019 LD2 are shown with their semimajor axes ('a', blue) and perihelion distances ('q', maroon). The clones are plotted at lower opacity. The clones are seen to diverge, especially in their semimajor axes, around the year ~1770, as seen by the marked spread in the semimajor axis evolution at earlier times. The perihelion distances remain clustered just outside of Jupiter's orbit (the heliocentric distance ranges of Jupiter and Saturn's orbits are indicated by the gray bands).

Over this well-determined history since ~1770, LD2's perihelion distance did not change dramatically, but its semimajor axis experienced two marked decreases (from a near-Saturn value to ~7-8 au around 1850 and then again to its current near-Jupiter value in 2017), producing a general increase in its average surface temperature throughout its orbit. Prior to ~1770, its semimajor axis is shown to diverge quite rapidly. While it is not possible to infer exactly how representative these longer backwards integrations are its exact history, even in a statistical sense (see Morbidelli et al. 2020), we do note that the vast majority of the clones maintain larger semimajor axes with perihelia clustered near Jupiter's orbit; relatively few of the clones enter the inner solar system (inside of ~4 au) in the past 3000 years (see Section 3.1).

From our understanding of the Centaur-to-JFC transition (Sarid et al. 2019), this distribution of possible dynamical histories combined with LD2's well-determined recent history and near-future is consistent with an inward transition from a near-Saturn Centaur orbit to a Gateway orbit between Jupiter and Saturn. This strongly suggests that LD2 has not previously been in the inner solar system, but is instead about to experience its first transition into the JFC population. As mentioned in Kareta et al. (2020) and Hsieh et al. (2020), LD2 is currently in-between two close approaches with Jupiter: one that occurred in 2017 at a distance of 0.092 au (0.272 Jovian Hill radii) that placed it into its current orbit, and an upcoming encounter in 2028 at a distance of 0.119 au (0.352 Jovian Hill radii) that will further change its orbit. In 2063, LD2 will have another close encounter with Jupiter at ~0.03-0.04 AU (< 0.1 Jovian Hill radii) that has a very high probability (>98%) of scattering it into the JFC population in the inner solar system.

### 3.1. Thermal History of 2019 LD2

Although JFCs are often thought to be the end result of a linear chain of transitions from TNO to Centaur and ultimately to JFC, these transitions are actually blurry; objects hop back and forth between populations. For example, Centaurs can follow Gateway orbits into the JFC population and back out again, complicating their thermal history. Gateway objects that previously spent significant time in the inner solar system (i.e., the JFC population) would likely be significantly dimmer and show much less activity than objects recently entering the Gateway from more distant regions of the Centaur population. Thus, a deeper understanding of an object's dynamical history is critical in understanding how to interpret the observed activity of Gateway objects.

To first order, the heat of the sun can thermally influence subsurface material within the thermal skin depth ($d_{skin}$) of the surface, which calculates the depth over which the thermal wave amplitude drops to 1/$e$ of its surface value in response to periodic heating with periodicity ($P$)

$$d_{skin} = \sqrt{HP} \qquad (1)$$

where $H$ is the thermal diffusivity (~$10^{-7}$-$10^{-8}$ for cometary bodies; Steckloff et al. 2020). This results in an orbital thermal skin depth of only ~2-5 m, for the current orbit. Thus, LD2's current variation in activity is likely due to variations within a few skin depths of the surface (~10-20 m). This relation can also be used to estimate the depth to which the thermal environment of the Gateway region can alter the nucleus near-surface material. The median residence in the Gateway is 1750 years, resulting in a thermal depth of ~20-70 m; even over the entire 3000 simulated years, this thermal depth increases to only ~30 - 100 m, due to this square-root dependence on time. Thus any material deeper than a few ~10's to ~100 m has yet to be affected by the inward evolution of LD2 into the Gateway's thermal environment.

However, LD2's evolving thermal environment could thermally alter materials within this thermal skin depth through solar heating. To understand the extent of thermal processing within the thermal skin depth, we compare the LD2's thermal environment as its orbit dynamically evolves with that of a "canonical" JFC orbit (semimajor axis of 3.5 au and eccentricity of 0.5; overall median values for known JFCs). In Figure 2 we compute the orbitally averaged surface insolation flux from the Sun as a function of orbital eccentricity and semimajor axis, and normalize by dividing by the orbitally averaged insolation flux of a canonical JFC. We then compare the evolving orbit of LD2 (all clones including the best-fit clone),with these normalized insolation fluxes.

We can see that, in general, LD2 has experienced less than half the solar energy input of JFCs, which has likely limited the extent of its subsurface thermal evolution. We find that, with the exception of a few of the clones, LD2's current orbit has the most intense thermal environment that this object has ever experienced. We also compare LD2's thermal environment with a few notable active object both inside and outside of the Gateway region.With the exception of Chiron, these active Centaurs are in orbits consistent with the dynamical evolution of LD2, suggesting that these objects may be experiencing a similar coupled dynamical and thermal evolution. In particular, 39P/Oterma has a rapidly evolving orbit, intermittent activity periods, and a high probability of previously residing in the JFC population, before migrating outward(Fernandez et al. 2001, Bauer et al. 2003, Toth 2006, Schambeau et al.2019).

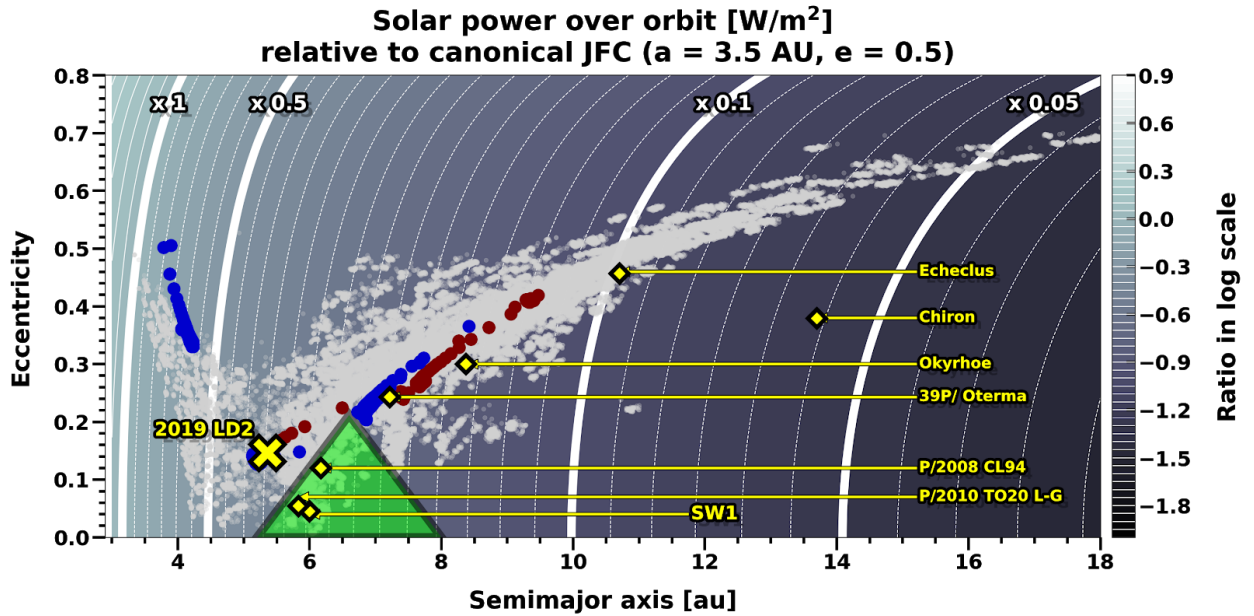

**Figure 2:** The recent eccentricity vs. semi-major axis evolution of LD2 shown in the context of its orbitally averaged thermal environment. Light grey dots represent the orbital evolution of 1000 clones of LD2 over the last 3000 years. Colorful dots represent the dynamical evolution of the best-fit clone, with red dots representing its past and blue dots representing its future evolution. These simulated evolutions are superimposed on a map of the average "heating intensity" (solar power per unit area, averaged over orbit) on an object's surface, which is normalized to the same value for a "canonical" JFC (see text). White contour curves represent the ratio between the calculated solar heating and that of a canonical JFC in log scale, with thick contour lines denoting 100%, 50%, 10%, and 5% of the heating of a canonical JFC. The yellow X marks the current orbit of LD2. This figure reveals that LD2 is being exposed to one of its most intense thermal environments in the past 3000 years. Yellow diamonds mark the current orbits of several notable active Centaurs, including SW1, the archetypal Gateway object (Sarid et al. 2019). The nominal Gateway region is marked with a green triangle.

To further constrain the probable recent history of LD2's thermal environment, we averaged the heliocentric distance bins of every clone in our backwards numerical integrations. We then integrated this average histogram to compute the expected fraction of the simulated time that it spends within or interior to a specified heliocentric bin[3]. Figure 3 shows the distribution of expectation values from this calculation (probability and cumulative distributions). However, these histograms are not to be interpreted as representing its actual history, which is only statistically known prior to ~1770 (see above). Indeed, individual clones may have divergent histograms, and spend significantly different amounts of time in each heliocentric bin. Rather, this is a computation of the expectation value of its likely residence duration in each heliocentric bin. We find that LD2 is likely to have spent the overwhelming majority (>90%) of the simulated time outside the orbit of Jupiter, and is unlikely to have spent significant amounts of time (~1.2 %) in the inner solar system (interior to 4 au), where water-ice sublimation begins to significantly influence cometary evolution (e.g., Womack et al. 2017). Furthermore, LD2 likely spent even less time (<0.25%) interior to 3 au, where there is a significant change in water's vapor pressure curve as water-ice sublimation becomes a comet's dominant cooling mechanism (e.g., Steckloff et al. 2015). Moreover, the majority of clones (~70%) spend no time inside 4 au; 84% spent no time inside of 3 au.

---

[3] These are effectively the statistical probability density function (PDF) and cumulative density function (CDF) of LD2's residence as a function of heliocentric distance.

This coupled dynamical and thermal history of LD2 supports an emerging view of an object that, despite showing considerable activity, is unlikely to have experienced thermal processing in the inner Solar System or significant bulk alteration in subsurface layers below a few skin depths. Indeed, LD2's subsurface is likely to have been processed by the Gateway thermal environment to a depth of no more than ~20-100 m; considering various estimates of nucleus size (see section 3.3) this is most likely a small fraction of LD2's volume. Thus, LD2 is primed to exhibit significant increased activity when it migrates closer to the Sun and is exposed to the more intense JFC thermal environment.

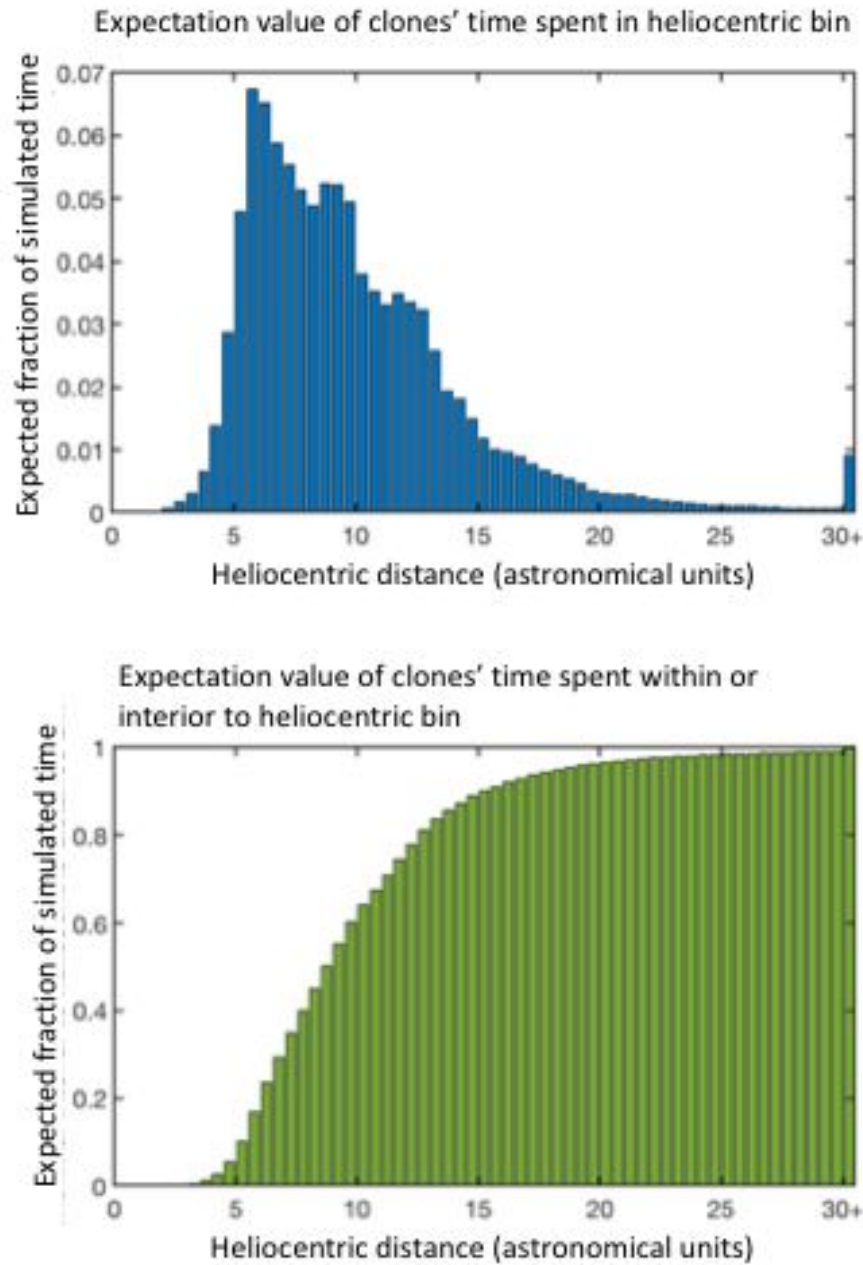

**Figure 3:** Plot of (*top*) probability density function (PDF) and (*bottom*) cumulative density function (CDF) of the ensemble of clones as a function of heliocentric distance over the past 3000 years. These plots show the expectation value of 2019 LD2's orbital evolution, showing (*top*) the expected fraction of the past 3000 years that 2019 LD2 spent in each heliocentric distance bin, and (*bottom*) the expected fraction of the past 3000 years that 2019 LD2 spent either within, or interior to, each heliocentric bin. These plots should not be interpreted as its actual dynamical evolution, as significant variations between the evolution of clones exist. Rather, these are statistical results that treat the ensemble of clones as equally probable.

### *3.2. Is 2019 LD2 a "Pristine" Object?*

An essential question regarding the scientific importance of LD2 is how pristine its nucleus is; to what degree of certainty can we determine whether it has spent significant time inside 4 au where water sublimation dominates evolution? Its orbital history, association with the Gateway region, and activity all point toward a past that most likely does not include time in the inner solar system. LD2's current orbit has a Jupiter Tisserand invariant of $T_J = 3.001$ and perihelion distance of q=4.58 au, which place it at the cusp of both the JFC dynamical class and the orbital domain where activity transitions to water sublimation (e.g. Bauer et al. 2015) and near the distance where CO outgassing is detected in other Centaurs (e.g., Womack et al. 2017; Wierzchos et al. 2017). Similarly, its orbital evolution is sufficiently deterministic to rule out any residency in the inner solar system since ~1770.

Earlier than ~1770, the orbital history of LD2 is chaotic and probabilistic. Our results show that, over the past 3000 years (since ~1000 BCE), it can be expected to have spent less than 1.2% of this time inside of 4 au, and a 70% chance of having never been inside of 4 au; similarly, it can be expected to have spent less than 0.25% of this time inside of 3 au, and an 84% chance of having never been inside of 3 au. Dynamical chaos beyond 1000 BCE dramatically reduces the statistical value of orbital models, but the nature of the Gateway itself provides an additional boundary condition. Centaurs entering the Gateway evolve rapidly due to their frequent encounters with Jupiter, with a median lifetime of ~1750 years or less before they transition to the JFCs or are lost (Sarid et al. 2019). The majority of Centaur objects that migrate into the Gateway region arrive from distant orbits (see e.g. Figure 2; Di Sisto & Brunini 2007; Sarid et al. 2019).

The above considerations of orbital evolution and activity behavior suggest that LD2 has not spent any significant time in the inner solar system during its entire evolution, as shown in Figure 2, prior to skidding through or just past the Gateway region in its near-future orbital evolution. Thus, its nucleus most likely has only lost volatiles that are active in the Centaur region while retaining those that drive JFC activity. Hence, it is likely to be a "pristine" object. In this context, "pristine" means that LD2's extent of surface and sub-surface thermophysical evolution has progressed to a much lesser extent, compared with the thermophysical processing experienced by comet nuclei in the inner solar system (i.e., the JFCs), its next dynamical residence population. Therefore, LD2 is the first object known that can be studied and observed to understand how this quintessential dynamical transition into the JFC population affects the activity, thermal evolution, and behavior of an icy body as it first enters and responds to the JFC thermal environment; thus, LD2 presents a uniquely compelling observational target over the coming decades.

### 3.3. Expected Number of 2019 LD2-Sized Objects in the Gateway

The four identified candidate objects in the Gateway (29P/Schwassmann-Wachmann 1, P/2010 TO20 (LINEAR-Grauer), P/2008 CL94 (Lemmon), and 2016 LN8) all have estimated radii between 2 and 6 km, except for Schwassmann-Wachmann 1, which has a radius of 23 to 32 km (Bauer et al. 2013; Schambeau et al. 2020a). While Schwassmann-Wachmann 1 is likely the only object of its size in the Gateway region (Sarid et al. 2019), it is presently unknown if this sample of smaller objects is complete, or how many LD2-sized objects should be in the Gateway.

We follow the procedure in Sarid et al. (2019) to make a simple estimate of how many LD2-sized objects should be present in Gateway orbits. LD2 has an absolute magnitude of $H_V =$ 12.2±0.8 (JPL Horizons). If we assume this value to represent the absolute magnitude of its nucleus, we can estimate the radius, by assigning an albedo value (see, e.g., Harris & Lagerros, 2002). Typical albedo values for Centaurs, range from 0.05 to 0.112 (Romanishin & Tegler 2018), resulting in LD2 radius estimates of $11^{+4.5}_{-3.5}$ km to $7^{+3.5}_{-2}$ km. However, absolute magnitudes for active objects like LD2 can incorporate brightness from the coma itself, so these nucleus sizes are most probably overestimates. DECam and Pan-STARRS 1 precovery data from 2018 observed LD2 with a low level of detectable activity (Schambeau et al. 2020b). These observations were used to estimate upper limits to LD2's nucleus radius: 5.0±0.1 km or 3.34±0.08 km for these two albedo values, respectively (Schambeau et al. 2020b).

Further confusing a size estimate of LD2, additional DECam precovery images from 2017 March and May did not detect LD2 within 10" of its ephemeris position (±0.2" 1-sigma ephemeris positional uncertainty), suggesting that its nucleus may be smaller still (Schambeau et al. 2020b). Ultimately, the size of LD2's nucleus is still poorly known, which prompts us to consider a range of applicable radii. This range spans values relevant to JFCs and other active bodies (e.g. Snodgrass et al. 2011) and estimated as: >1 km, >3 km, >5 km, and >10 km.

Assuming the Centaur population follows a power-law size distribution
$$dN_r = -k\alpha r^{-(\alpha+1)} dr \qquad (2)$$
where $\alpha = 3$ and $k = 6.5 \times 10^6 \ km^{-1}$ (Sarid et al. 2019), we estimate that there are: ~6.5 million Centaurs with radius >1 km, ~240,000 Centaurs >3 km, ~52,000 Centaurs >5 km, and ~6,500 Centaurs >10 km. From the typical duration of a Centaur's residency in a Gateway orbit relative to its residency in the Centaur population as a whole, and the knowledge that 21% of Centaurs have a Gateway phase (Sarid et al. 2019), we can estimate the current Gateway population as a function of radius. Considering only the gravitational evolution of Centaurs, we estimate that the Gateway population currently contains ~1,000 Centaurs with radius >1 km, ~37 Centaurs >3 km, ~8 Centaurs >5 km, and ~1 Centaur >10 km. If we use the Brasser & Wang (2015) empirical fading law to account for processes in the inner solar system that remove objects from the population, we estimate that the Gateway population currently has ~240 Centaurs >1 km, ~9 Centaurs >3 km, ~2 Centaurs >5 km, and ~0.24 Centaur >10 km in radius. This suggests that the population of Gateway objects larger than 3 km, is partially complete. However uncertainties in LD2's size limit our ability to understand how complete its size-class is; additional observations are therefore required to constrain this fundamental property.

| Centaur radius | Number of Centaurs in Gateway | | Median Frequency transitions: Gateway → JFC |
|---|---|---|---|
| | (no fading) | (fading) | |
| >1 km | ~1,000 | ~240 | ~2.7 years |
| >3 km | ~37 | ~9 | ~73 years |
| >5 km | ~8 | ~2 | ~340 years |
| >10 km | ~1 | ~0.24 | ~2,700 years |

**Table 1:** Number of Centaurs and Gateway Objects, and frequency of Transition from Gateway to JFC population as a function of radius. Size ranges are specified as minimum size and up; size-frequency distribution is heavily weighted toward smaller sizes. The "no fading" column reflects the expected number of objects in the Gateway due purely to gravitational considerations. The "fading" column uses the Brasser & Wang (2015) empirical fading law, as implemented in Sarid et al. (2019) to remove objects that have spent sufficient time in the inner solar system to sublimate away, disrupt, or become otherwise removed from the population. Ultimately, fading/no fading does not affect the frequency of transitions, only the number of objects in the Gateway.

The median duration of residency in Gateway orbits is 1750 years without fading, and 425 years with a fading law, with ~65% of Gateway objects ultimately transitioning into the JFC population (Sarid et al. 2019). If LD2 is in the smallest size range (>1 km radius), such objects transition out of the Gateway and into the JFC population with a median frequency of once every ~2.7 years. This is the size-range of nearly every known JFC (Snodgrass et al. 2011; Fernández et al. 2013); in this case, LD2's migration from Centaur to the JFC population in 2063 (Kareta et al. 2020) represents a unique opportunity to observe how a typical object responds to the changing thermal environment associated with this transition. However, if LD2 is larger (greater than 3 km in radius), such transitions become increasingly rare, occurring with a median frequency of no more than once every ~73 years. In this case, its transition into the JFC population would represent a once-in-a-lifetime opportunity to study how a potentially Great Comet would respond to this transition. In either case, LD2's upcoming transition into the JFC population provides the first known opportunity to observe how an active, pristine Centaur responds to this transition into the JFCs. That this transition will occur within our lifetimes makes LD2 a uniquely compelling object to study, and suggests that follow up observations, along with a long-term monitoring campaign are highly likely to produce important, scientifically impactful results.

### 4. Conclusions

We used a series of 1,000 clones to study 2019 LD2's dynamical evolution statistically over the past 3000 years. Our dynamical simulations find that its orbit can be deterministically integrated backward in time until ~1770, beyond which point further integration becomes chaotic. We find that LD2 is currently passing through the "dynamical Gateway" that facilitates most transitions between the Centaurs and JFCs (Sarid et al. 2019); showing the importance of the Gateway to the dynamical evolution of comets. Furthermore, we find that LD2 is unlikely to have spent significant time in the inner solar system, and thus unlikely to have experienced significant thermophysical evolution of its nucleus. We conclude that LD2 is likely to be a relatively pristine nucleus that has only been affected by minor thermal processing in the Centaur region (e.g. Sarid & Prialnik 2009). This is consistent with its observed high level of activity for a Centaur at its current heliocentric distance of 4.6 au.

Finally, we use our own and previous dynamical studies and a size frequency distribution for the Centaurs to estimate how many LD2-like objects are currently in the Gateway region. We conclude that there is a steady-state group of LD2-like objects (Gateway objects imminently transitioning into the JFC population); the frequency of these transitions depends sensitively on the radius of its nucleus. An LD2-like object the size of a typical JFC transitions into the JFC population with a median frequency of once every ~2.7 years; if LD2 is this size, its transition into the JFC population in 2063 presents the first known opportunity to observe how a typical, pristine JFC's activity, appearance, and dynamical behavior evolve as it progresses in its orbital migration. However, an LD2-like object ~3 km, ~5km, and ~10km or larger in radius transitions, respectively, into the JFC population once every ~73, ~340, and ~2,700 years. Thus, if it is larger than a typical JFC, its imminent transition also presents a once-in-a-lifetime opportunity to observe how a large, pristine, and potentially Great Comet responds to the changing thermal environment associated with this defining transition.

P/2019 LD2 (ATLAS) is the first active Centaur caught in mid-transition to the JFC population, and its level of activity suggests it is in a near-pristine thermal state. This underscores the need for more observational and theoretical studies of this object, and the Gateway region, where we can search for more objects going through similar transitions (Sarid et al. 2019). Specifically, our understanding of both P/2019 LD2 (ATLAS) and the wider Gateway population would benefit greatly from dedicated exploration endeavours that can uncover the nature of this quintessential evolution from Centaur to JFC, through survey observations, long term monitoring, and in-situ measurements.

## 5. Acknowledgements

We wish to thank the anonymous reviewer, whose comments improved and strengthened this work. J.K.S. acknowledges support from NSF grant 1910275 and NASA award 80NSSC19K1313. G. S acknowledges support from NASA award 80NSSC18K0497. This material is based on work supported by the National Science Foundation under Grant No. AST-1945950 (MW as PI).